\documentclass{iaus}
\usepackage{graphicx}
\usepackage{natbib}

\title[Modelling stellar coronal magnetic fields] 
{Modelling stellar coronal magnetic fields}

\author[M. Jardine, J.-F. Donati]   
{Moira Jardine$^1$, Jean-Francois Donati$^2$,  Doris Arzoumanian$^1$ and Aline de Vidotto$^1$}

\affiliation{$^1$SUPA, School of Physics and Astronomy, University of St Andrews, North Haugh, St Andrews, KY16 9SS, UK \\ email: {\tt mmj@st-andrews.ac.uk,Aline.Vidotto@st-andrews.ac.uk,doris.arzoumanian@cea.fr} \\[\affilskip]
$^2$LATT, CNRS--UMR 5572, Obs.\ Midi-Pyr\'en\'ees, 14 Av.\ E.~Belin, F--31400 Toulouse, France \\email: {\tt donati@ast.obs-mip.fr}} 

\pubyear{2010}
\volume{273}  
\pagerange{***--***}
\jname{Physics of Sun and Star Spots}
\editors{A.C. Editor, B.D. Editor \& C.E. Editor, eds.}
\begin{document}

\maketitle

\begin{abstract}
Our understanding of the structure and dynamics of stellar coronae has changed dramatically with the availability of surface maps of both star spots and also magnetic field vectors. Magnetic field extrapolations from these surface maps reveal surprising coronal structures for stars whose masses and hence internal structures and dynamo modes may be very different from that of the Sun. Crucial factors are the fraction of open magnetic flux (which determines the spin-down rate for the star as it ages) and the location and plasma density of closed-field regions, which determine the X-ray and radio emission properties. There has been recent progress in modelling stellar coronae, in particular the relative contributions of the field detected in the bright surface regions and the field that may be hidden in the dark star spots. For the Sun, the relationship between the field in the spots and the large scale field is well studied over the solar cycle. It appears, however, that other stars can show a very different relationship.

\end{abstract}

\firstsection 
\section{Introduction}

Just as the surface distributions of spots on stars can vary considerably between different types of stars, so too can the structure of their coronae. For massive stars, the nature of the coronal magnetic field  may have an influence on the structure of the wind (which can remove a significant fraction of the star's mass over its lifetime). We do not, however, know if the magnetic fields of these stars are generated by dynamo processes \citep{charbonneau_macgregor_dynamo_01,brun_dynamo_05,spruit_dynamo_02,tout_pringle_dynamo_95,macdonald_mullan_dynamo_04,mullan_macdonald_dynamo_05,maeder_meynet_dynamo_05} or if they are fossil fields \citep{moss_review_01,braithwaite_spruit_fossilfields_04,braithwaite_nordlund_dynamo_06}. An equally challenging question is the source and generation mechanism of the observed X-ray emission \citep{ignace_tausco_10}. 

In the case of solar-mass stars, the nature of the magnetic cycles and the lifetimes of active regions are topics that will benefit enormously from the data that will soon be available from CoRoT and Kepler. As has been mention already in this symposium, the variation with mass of the differential rotation is another crucial question. At the bottom end of the main sequence, the low mass, fully convective stars appear to have much simpler, stronger fields than their higher-mass counterparts. While a decade or so ago, it was believed that these stars could only generate small-scale magnetic fields  \citep{durney_turb_dynamo_93, cattaneo_dynamo_99}, more recent studies have suggested that large scale fields may be generated. These models differ, however, in their predictions for the form of this field and the associated latitudinal differential rotation. They predict that the fields should be either axisymmetric with pronounced differential rotation \citep{dobler_dynamos_06}, non-axisymmtric with minimal differential rotation \citep{kuker_rudiger_97,kuker_rudiger_99,chabrier_kuker_06}. More recently, however, \citet{browning_dynamo_08} has published a dynamo model that produces a highly symmetric field with little differential rotation. This field structure may have implications for the nature of the spin-down of these low mass stars, since they show puzzlingly-high rotation rates, even on the main sequence. 
\begin{figure*}[t]
\begin{center}

 \includegraphics[width=4.in]{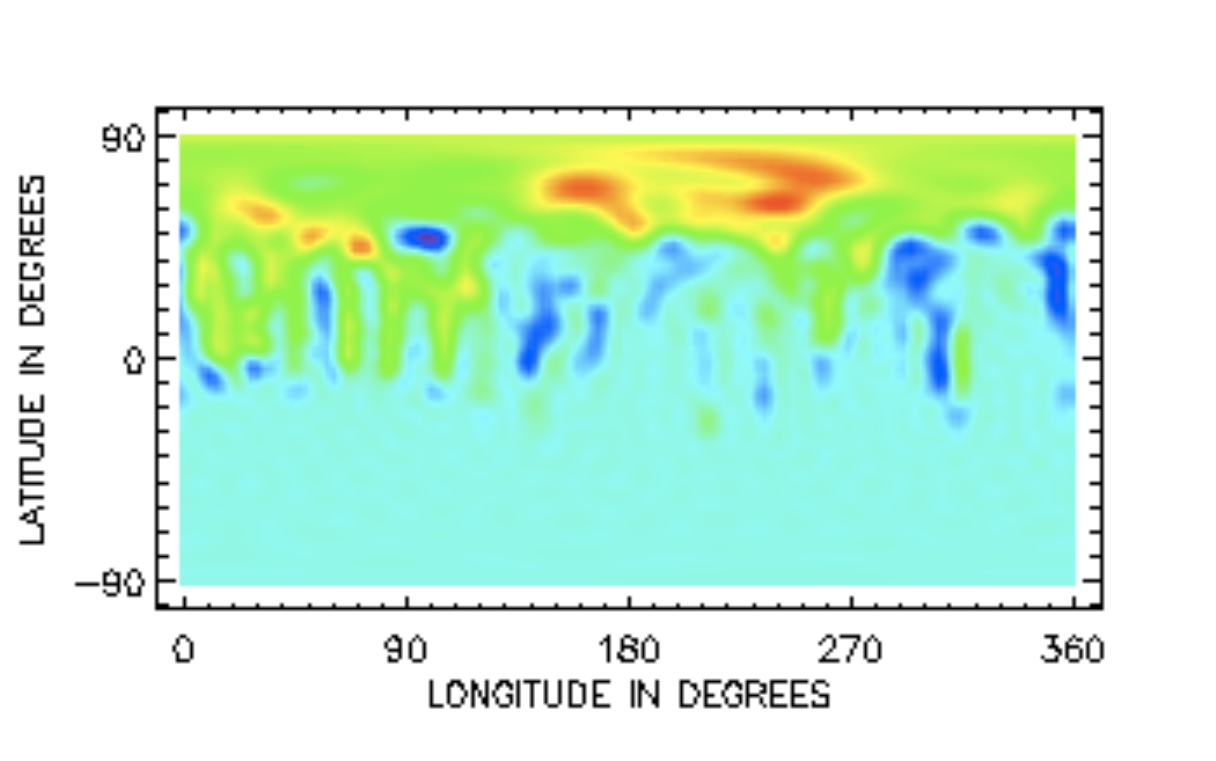} 
 \includegraphics[width=4.5in]{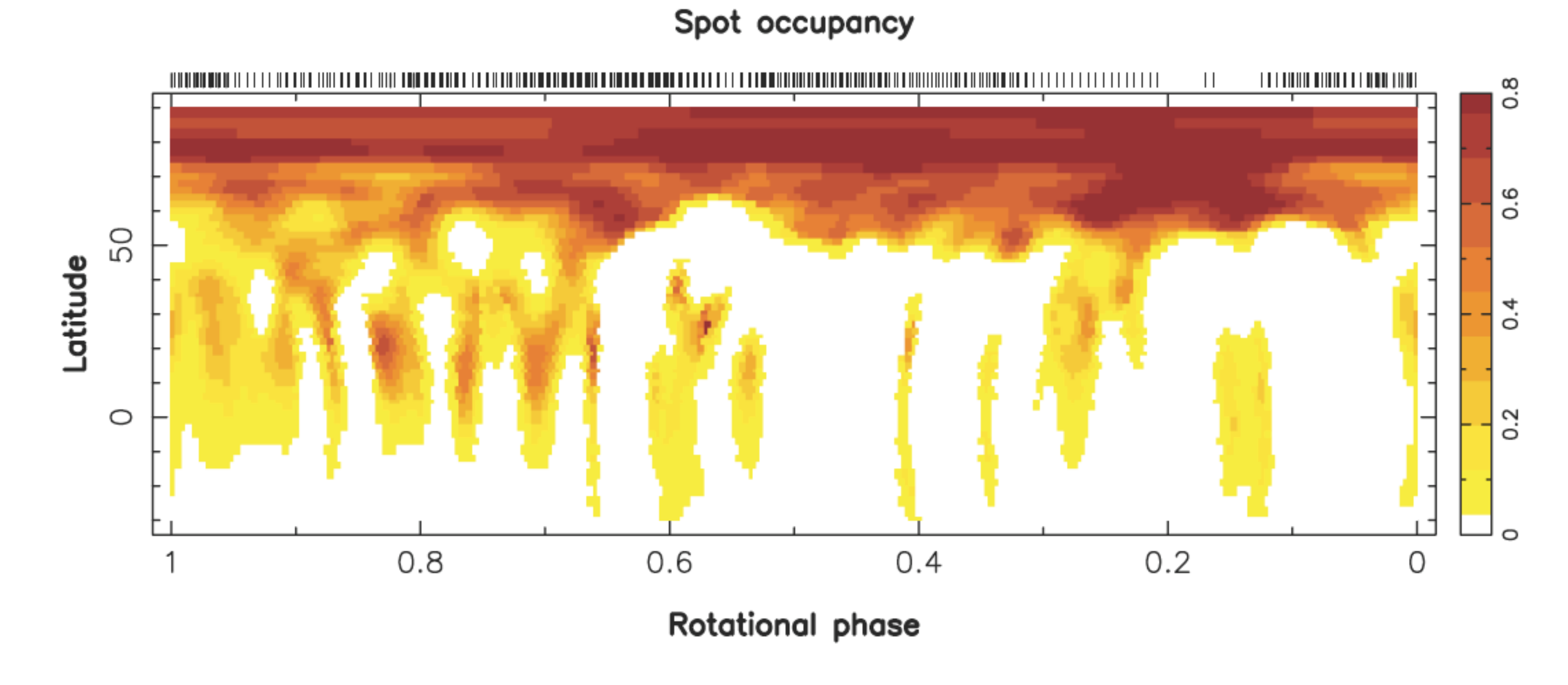} 

 \caption{Zeeman-Doppler maps of the radial field component (top) and the corresponding spot occupancy map (bottom) for the AB Dor (observations carried out in  2004).}
   \label{fig1}
\end{center}
\end{figure*}

Even in the pre-main sequence phase, the structure of  star's corona can be important since it affects the rate at which the star can accrete mass from its disk \citep{gregory_mdot_06,gregory_review_10}. Low mass stars remain fully convective throughout their evolution onto the main sequence, but higher mass stars will develop a radiative core at some stage - perhaps even while they are still accreting from their disks. This difference in internal structure may lead to different dynamo behaviours, and indeed there is a suggestion from observations of a handful of stars that the field structures of stars with a radiative core (and hence a tachocline where a solar-like interface dynamo might be active) are more complex than those of the fully-convective stars \citep{donati_v2129oph_07,donati_bptau_08,jardine_v2129oph_08,gregory_bptau_v2129oph_08,hussain_chacha_09}. The implications for the spin evolution of the stars have not yet been fully explored however.

\section{Modelling stellar coronae}
The first step in modelling the corona of a star is to determine the form of the surface field. This is most commonly done using the technique of Zeeman-Doppler imaging  \citep{donati97abdor95,donati99abdor96}. This typically shows a complex distribution of surface spots that is often very different from that of the Sun, with spots and mixed polarity flux elements extending over all latitudes up to the pole.  A comprehensive review of starspot distributions is given in \citet{strassmeier_review_09}. 
\begin{figure*}
\begin{center}

\includegraphics[width=2.5in]{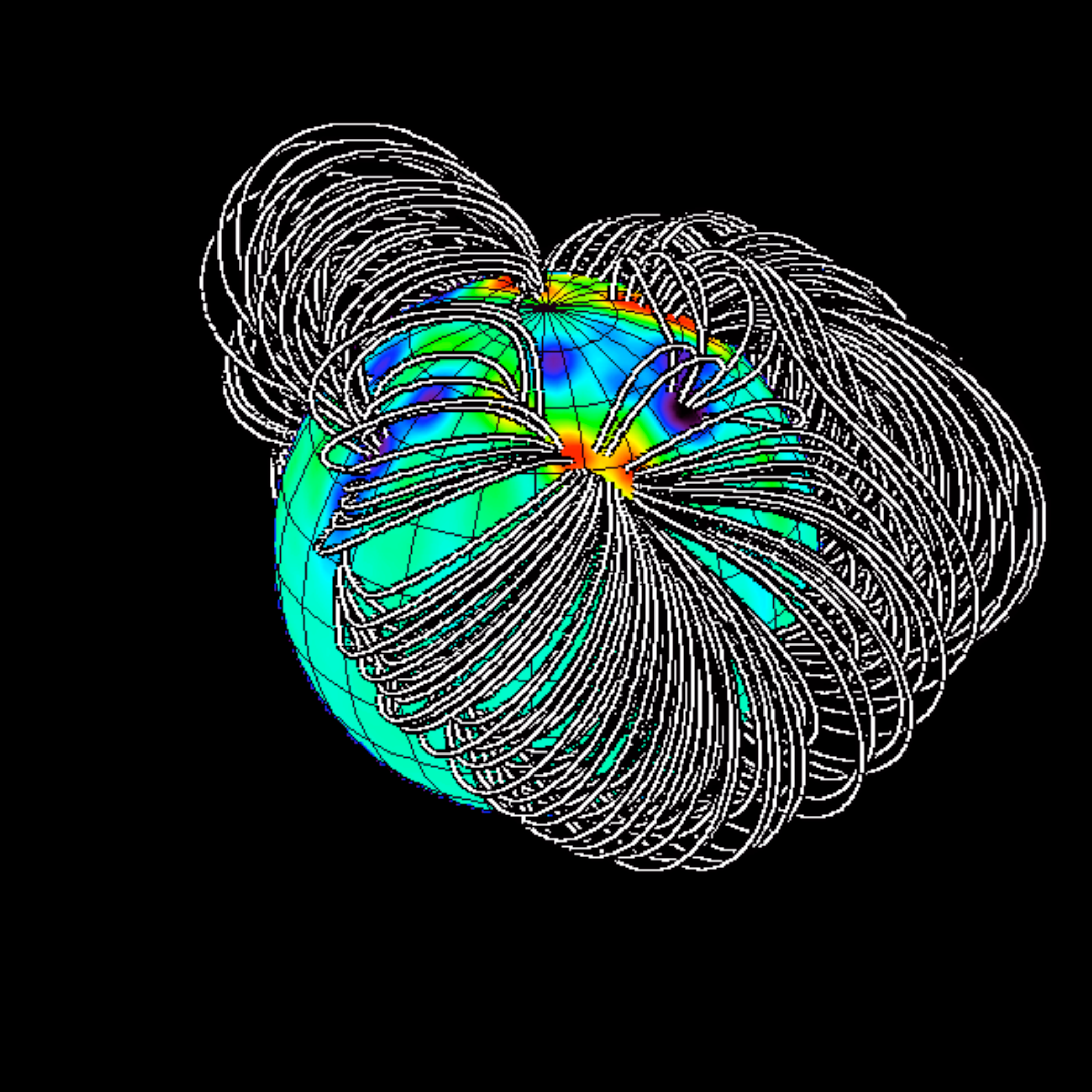}
\includegraphics[width=2.5in]{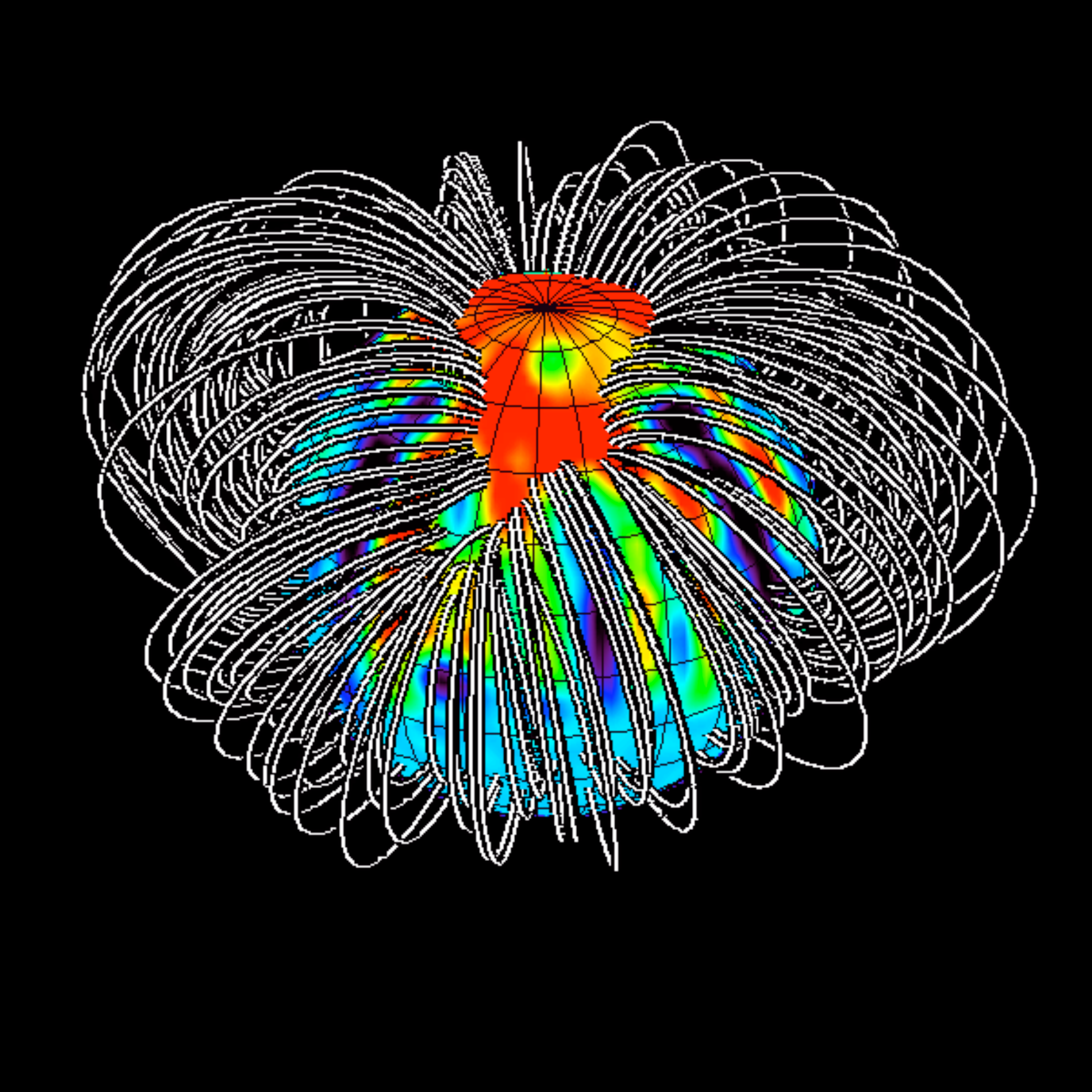}

 \caption{Field extrapolation from the original Zeeman-Doppler map shown in Fig. \ref{fig1} (left) and from the maps of the combined Zeeman-Doppler and spot field (right).}

  \label{fig2}
 \end{center}
\end{figure*}

While the methods of extrapolating the magnetic field from these magnetograms may differ, the basic procedure is the same. The most commonly-used method is the {\it Potential Field Source Surface} method \citep{altschuler69,jardine99ccf,jardine2001eqm,jardine02structure,mcivor03polar}, but it is also possible to use non-potential fields \citep{donati01,hussain02nonpot}. Once the 3D structure of the magnetic field has been determined, the structure of the coronal gas can be determined by assuming that the gas trapped on these field lines is in isothermal, hydrostatic equilibrium. We can then determine the coronal gas pressure, subject to an assumption for the gas pressure at the base of the corona. We assume that it is proportional to the magnetic pressure, i.e. $p_0 \propto B_0^2$, where the constant of proportionality is determined by comparison with X-ray emission measures \citep{jardine02xray,jardine_TTS_06,gregory_rotmod_06}. For an optically thin coronal plasma, this then allows us to produce images of the X-ray emission.
\begin{figure*}
\begin{center}

\includegraphics[width=5.in]{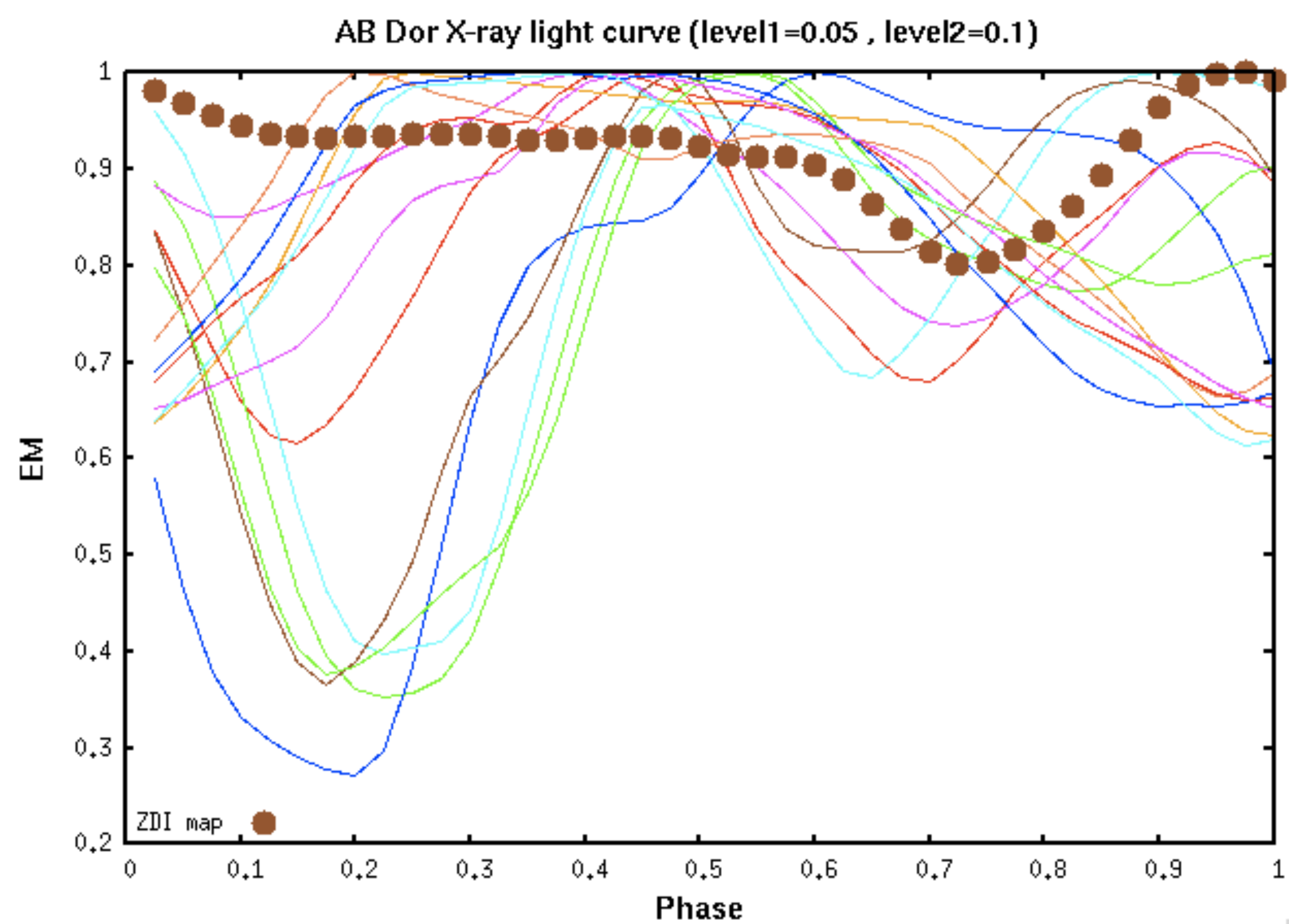}

 \caption{X-ray rotational modulation of AB Dor. The dots represent the X-ray rotational modulation calculated from the  ZDI map. The lines represent the  X-ray rotational modulation  using  the combined maps (ZDI + spot maps.}

  \label{fig3}
 \end{center}
\end{figure*}

\section{Effect of spot magnetic fields on coronal structure}

The fraction of a star's surface that is covered in spots and the strength and degree of complexity of the coronal magnetic field all vary with stellar mass. While low mass, fully convective stars commonly have strong, simple surface fields with few spots \citep{morin_earlyM_08,morin_midM_08,morin_lateM_10}, solar mass stars such as AB Dor (shown in Fig \ref{fig1}) typically have complex fields and spots that can cover a significant fraction of their surfaces \citep{strassmeier_review_09}. Zeeman-Doppler imaging does not recover all of the magnetic field in these spotted regions - a factor that may be more important in stars with greater spot coverage \citep{johnstone_spots_10}. Although we do no know the polarity of the magnetic field in the spots, we can take a statistical approach. We randomly allocate a polarity to each spot (defined at a certain level of spot occupancy) and then add this field to the surface field reconstructed by Zeeman-Doppler imaging. By repeating this process many times  and each time extrapolating the coronal field, we can create many realisations of the structure of the corona (see for example Fig.s \ref{fig2}). By doing this for stars of different masses, we can determine the possible effect that the flux hidden in the spots might have on the corona \citep{arzoumanian_10}. Because the spot field typically has a greater degree of axial symmetry, this process often shifts the large-scale dipole axis closer to the rotation axis. The fraction of open flux and the magnitude of the X-ray emission measure also typically increase. Fig. \ref{fig3} shows some examples of the X-ray light curves that might be produced.

\section{Moving towards full MHD}

\begin{figure}[t]
\begin{center}

    \includegraphics[width=4.in]{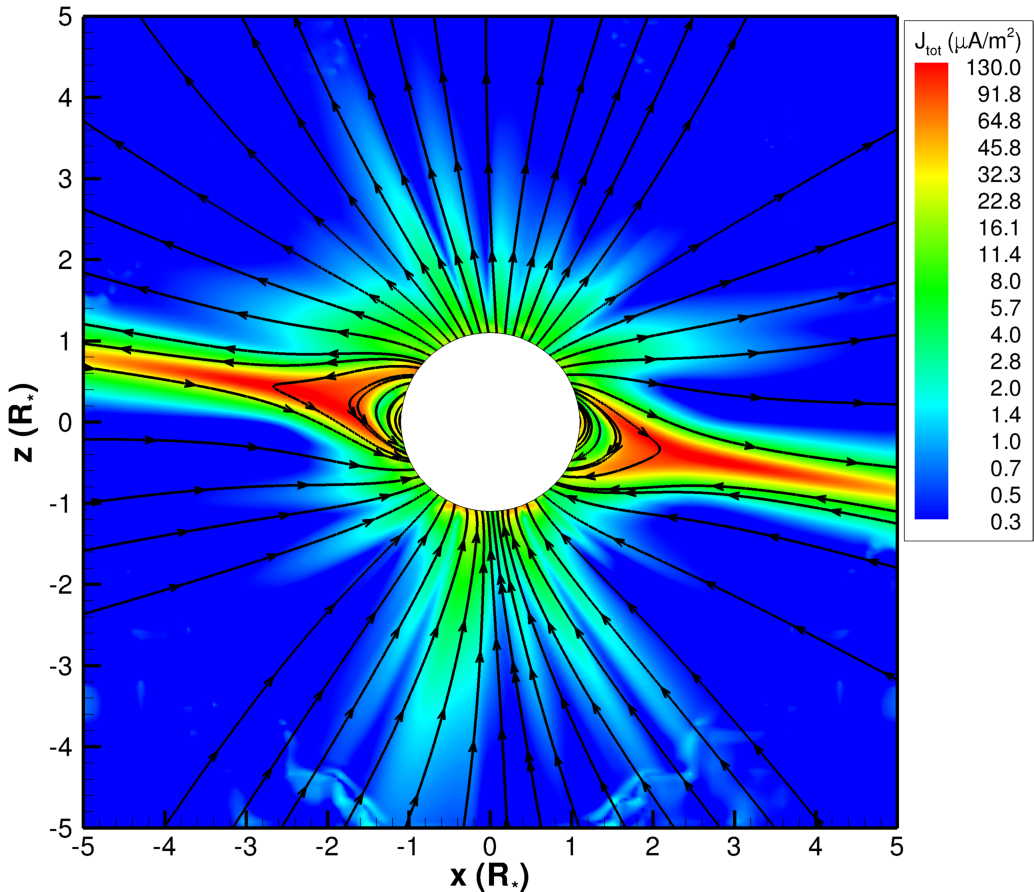} 

 \caption{Current density.}
   \label{fig4}
\end{center}
\end{figure}

While the potential field source surface models have the advantage of computational speed and simplicity, they only allow indirectly for the effect of the stellar wind. In order to improve our understanding of the interplay between coronal and wind processes in stars we need to incorporate the observed surface magnetograms into an MHD code \citep{cohen_abdor_10, vidotto_v374Peg_10}. First results from this for the fully-convective star V374 Peg have been extremely interesting (see Fig. \ref{fig4}). Like many low mass stars, V374 Peg rotates extremely rapidly and has a strong, almost dipolar field. This means that magnetic forces dominate over the other forces and results in a very non-solar, magneto-centrifugally driven wind. The general scalings that emerge from this study are that both the angular momentum loss rate and the mass loss rate scale as n$^{1/2}$ (where n is the density) and hence the spin-down time scales as n$^{-1/2}$. The fact that V374 Peg is a rapid rotator therefore suggests it has a coronal density that is low by stellar standards, if high relative to the Sun. The mass loss rate and speed of the wind are both also higher than solar values. Most interesting, however, is the ram pressure which may be five orders of magnitude higher than that of the solar wind. This has significant implications for planet habitability, since such a high ram pressure may crush planetary magnetospheres. If, for example, we place a planet in the habitable zone of an M-dwarf star, it needs a field of about 8G (around half that of Jupiter) to maintain a magnetosphere. The means by which such a field might be generated in the planetary interior are still a matter of debate.


\section{Conclusions}

The availability of spot maps and magnetograms for stars other than the Sun has revolutionised our view of stellar magnetic fields. By using these maps to model the coronal structure of a range of stars we can also gain insight into their winds, their rotational evolution and their coronal emission. Advances in MHD modelling that incorporate the three vector components of the surface magnetic field promise to bring about a similar revolution.


\begin{thebibliography}{45}
\expandafter\ifx\csname natexlab\endcsname\relax\def\natexlab#1{#1}\fi

\bibitem[{{Altschuler} \& {Newkirk, Jr.}(1969)}]{altschuler69}
{Altschuler}, M.~D. \& {Newkirk, Jr.}, G. 1969, Solar~Phys., 9, 131

\bibitem[{{Arzoumanian} {et~al.}(2010){Arzoumanian}, {Jardine}, {Donati},
  {Morin}, \& {Johnstone}}]{arzoumanian_10}
{Arzoumanian}, D., {Jardine}, M., {Donati}, J., {Morin}, J., \& {Johnstone}, C.
  2010, ArXiv e-prints

\bibitem[{{Braithwaite} \& {Nordlund}(2006)}]{braithwaite_nordlund_dynamo_06}
{Braithwaite}, J. \& {Nordlund}, A. 2006, A\&A, 450, 1077

\bibitem[{{Braithwaite} \& {Spruit}(2004)}]{braithwaite_spruit_fossilfields_04}
{Braithwaite}, J. \& {Spruit}, H.~C. 2004, Nature, 431, 819

\bibitem[{{Browning}(2008)}]{browning_dynamo_08}
{Browning}, M.~K. 2008, ApJ, 676, 1262

\bibitem[{{Brun} {et~al.}(2005){Brun}, {Browning}, \&
  {Toomre}}]{brun_dynamo_05}
{Brun}, A.~S., {Browning}, M.~K., \& {Toomre}, J. 2005, ApJ, 629, 461

\bibitem[{{Cattaneo}(1999)}]{cattaneo_dynamo_99}
{Cattaneo}, F. 1999, ApJ, 515, L39

\bibitem[{{Chabrier} \& {K\"uker}(2006)}]{chabrier_kuker_06}
{Chabrier}, G. \& {K\"uker}, M. 2006, A\&A, 446, 1027

\bibitem[{{Charbonneau} \& {MacGregor}(2001)}]{charbonneau_macgregor_dynamo_01}
{Charbonneau}, P. \& {MacGregor}, K.~B. 2001, ApJ, 559, 1094

\bibitem[{{Cohen} {et~al.}(2010){Cohen}, {Drake}, {Kashyap}, {Hussain}, \&
  {Gombosi}}]{cohen_abdor_10}
{Cohen}, O., {Drake}, J.~J., {Kashyap}, V.~L., {Hussain}, G.~A.~J., \&
  {Gombosi}, T.~I. 2010, ArXiv e-prints

\bibitem[{{Dobler} {et~al.}(2006){Dobler}, {Stix}, \&
  {Brandenburg}}]{dobler_dynamos_06}
{Dobler}, W., {Stix}, M., \& {Brandenburg}, A. 2006, ApJ, 638, 336

\bibitem[{{Donati} {et~al.}(2008{\natexlab{a}}){Donati}, {Morin}, {Petit},
  {Delfosse}, {Forveille}, {Auri{\`e}re}, {Cabanac}, {Dintrans}, {Fares},
  {Gastine}, {Jardine}, {Ligni{\`e}res}, {Paletou}, {Velez}, \&
  {Th{\'e}ado}}]{morin_earlyM_08}
{Donati}, J., {Morin}, J., {Petit}, P., {et~al.} 2008{\natexlab{a}}, MNRAS,
  390, 545

\bibitem[{{Donati}(2001)}]{donati01}
{Donati}, J.-F. 2001, LNP Vol.~573: Astrotomography, Indirect Imaging Methods
  in Observational Astronomy, 573, 207

\bibitem[{Donati \& Collier~Cameron(1997)}]{donati97abdor95}
Donati, J.-F. \& Collier~Cameron, A. 1997, MNRAS, 291, 1

\bibitem[{Donati {et~al.}(1999)Donati, Collier~Cameron, Hussain, \&
  Semel}]{donati99abdor96}
Donati, J.-F., Collier~Cameron, A., Hussain, G., \& Semel, M. 1999, MNRAS, 302,
  437

\bibitem[{{Donati} {et~al.}(2007){Donati}, {Jardine}, {Gregory}, {Petit},
  {Bouvier}, {Dougados}, {M{\'e}nard}, {Cameron}, {Harries}, {Jeffers}, \&
  {Paletou}}]{donati_v2129oph_07}
{Donati}, J.-F., {Jardine}, M.~M., {Gregory}, S.~G., {et~al.} 2007, MNRAS, 380,
  1297

\bibitem[{{Donati} {et~al.}(2008{\natexlab{b}}){Donati}, {Jardine}, {Gregory},
  {Petit}, {Paletou}, {Bouvier}, {Dougados}, {M{\'e}nard}, {Cameron},
  {Harries}, {Hussain}, {Unruh}, {Morin}, {Marsden}, {Manset}, {Auri{\`e}re},
  {Catala}, \& {Alecian}}]{donati_bptau_08}
{Donati}, J.-F., {Jardine}, M.~M., {Gregory}, S.~G., {et~al.}
  2008{\natexlab{b}}, MNRAS, 386, 1234

\bibitem[{{Durney} {et~al.}(1993){Durney}, {De Young}, \&
  {Roxburgh}}]{durney_turb_dynamo_93}
{Durney}, B.~R., {De Young}, D.~S., \& {Roxburgh}, I.~W. 1993, Solar~Phys.,
  145, 207

\bibitem[{{Gregory} {et~al.}(2006{\natexlab{a}}){Gregory}, {Jardine},
  {Cameron}, \& {Donati}}]{gregory_rotmod_06}
{Gregory}, S.~G., {Jardine}, M., {Cameron}, A.~C., \& {Donati}, J.-F.
  2006{\natexlab{a}}, MNRAS, 373, 827

\bibitem[{{Gregory} {et~al.}(2010){Gregory}, {Jardine}, {Gray}, \&
  {Donati}}]{gregory_review_10}
{Gregory}, S.~G., {Jardine}, M., {Gray}, C.~G., \& {Donati}, J. 2010, ArXiv
  e-prints

\bibitem[{{Gregory} {et~al.}(2006{\natexlab{b}}){Gregory}, {Jardine},
  {Simpson}, \& {Donati}}]{gregory_mdot_06}
{Gregory}, S.~G., {Jardine}, M., {Simpson}, I., \& {Donati}, J.-F.
  2006{\natexlab{b}}, MNRAS, 371, 999

\bibitem[{{Gregory} {et~al.}(2008){Gregory}, {Matt}, {Donati}, \&
  {Jardine}}]{gregory_bptau_v2129oph_08}
{Gregory}, S.~G., {Matt}, S.~P., {Donati}, J., \& {Jardine}, M. 2008, MNRAS,
  389, 1839

\bibitem[{{Hussain} {et~al.}(2009){Hussain}, {Collier Cameron}, {Jardine},
  {Dunstone}, {Ramirez Velez}, {Stempels}, {Donati}, {Semel}, {Aulanier},
  {Harries}, {Bouvier}, {Dougados}, {Ferreira}, {Carter}, \&
  {Lawson}}]{hussain_chacha_09}
{Hussain}, G.~A.~J., {Collier Cameron}, A., {Jardine}, M.~M., {et~al.} 2009,
  MNRAS, 398, 189

\bibitem[{{Hussain} {et~al.}(2002){Hussain}, {van Ballegooijen}, {Jardine}, \&
  {Collier Cameron}}]{hussain02nonpot}
{Hussain}, G.~A.~J., {van Ballegooijen}, A.~A., {Jardine}, M., \& {Collier
  Cameron}, A. 2002, ApJ, 575, 1078

\bibitem[{{Ignace} {et~al.}(2010){Ignace}, {Oskinova}, {Jardine}, {Cassinelli},
  {Cohen}, {Donati}, {Townsend}, \& {ud-Doula}}]{ignace_tausco_10}
{Ignace}, R., {Oskinova}, L.~M., {Jardine}, M., {et~al.} 2010, ArXiv e-prints

\bibitem[{{Jardine} {et~al.}(1999){Jardine}, {Barnes}, {Donati}, \& {Collier
  Cameron}}]{jardine99ccf}
{Jardine}, M., {Barnes}, J., {Donati}, J.-F., \& {Collier Cameron}, A. 1999,
  MNRAS, 305, L35

\bibitem[{{Jardine} {et~al.}(2002{\natexlab{a}}){Jardine}, {Collier Cameron},
  \& {Donati}}]{jardine02structure}
{Jardine}, M., {Collier Cameron}, A., \& {Donati}, J.-F. 2002{\natexlab{a}},
  MNRAS, 333, 339

\bibitem[{{Jardine} {et~al.}(2006){Jardine}, {Collier Cameron}, {Donati},
  {Gregory}, \& {Wood}}]{jardine_TTS_06}
{Jardine}, M., {Collier Cameron}, A., {Donati}, J.-F., {Gregory}, S.~G., \&
  {Wood}, K. 2006, MNRAS, 367, 917

\bibitem[{{Jardine} {et~al.}(2001){Jardine}, {Collier Cameron}, {Donati}, \&
  {Pointer}}]{jardine2001eqm}
{Jardine}, M., {Collier Cameron}, A., {Donati}, J.-F., \& {Pointer}, G. 2001,
  MNRAS, 324, 201

\bibitem[{{Jardine} {et~al.}(2002{\natexlab{b}}){Jardine}, {Wood}, {Collier
  Cameron}, {Donati}, \& {Mackay}}]{jardine02xray}
{Jardine}, M., {Wood}, K., {Collier Cameron}, A., {Donati}, J.-F., \& {Mackay},
  D.~H. 2002{\natexlab{b}}, MNRAS, 336, 1364

\bibitem[{{Jardine} {et~al.}(2008){Jardine}, {Gregory}, \&
  {Donati}}]{jardine_v2129oph_08}
{Jardine}, M.~M., {Gregory}, S.~G., \& {Donati}, J. 2008, MNRAS, 386, 688

\bibitem[{{Johnstone} {et~al.}(2010){Johnstone}, {Jardine}, \&
  {Mackay}}]{johnstone_spots_10}
{Johnstone}, C., {Jardine}, M., \& {Mackay}, D.~H. 2010, MNRAS, 404, 101

\bibitem[{{K\"uker} \& {R\"udiger}(1997)}]{kuker_rudiger_97}
{K\"uker}, M. \& {R\"udiger}, G. 1997, A\&A, 328, 253

\bibitem[{{K\"uker} \& {R\"udiger}(1999)}]{kuker_rudiger_99}
{K\"uker}, M. \& {R\"udiger}, G. 1999, in ASP Conf. Ser. 178: Workshop on
  stellar dynamos, Vol. 178, 87--96

\bibitem[{{MacDonald} \& {Mullan}(2004)}]{macdonald_mullan_dynamo_04}
{MacDonald}, J. \& {Mullan}, D.~J. 2004, MNRAS, 348, 702

\bibitem[{{Maeder} \& {Meynet}(2005)}]{maeder_meynet_dynamo_05}
{Maeder}, A. \& {Meynet}, G. 2005, A\&A, 440, 1041

\bibitem[{{McIvor} {et~al.}(2003){McIvor}, {Jardine}, {Cameron}, {Wood}, \&
  {Donati}}]{mcivor03polar}
{McIvor}, T., {Jardine}, M., {Cameron}, A.~C., {Wood}, K., \& {Donati}, J.-F.
  2003, MNRAS, 345, 601

\bibitem[{{Morin} {et~al.}(2008){Morin}, {Donati}, {Petit}, {Delfosse},
  {Forveille}, {Albert}, {Auri{\`e}re}, {Cabanac}, {Dintrans}, {Fares},
  {Gastine}, {Jardine}, {Ligni{\`e}res}, {Paletou}, {Ramirez Velez}, \&
  {Th{\'e}ado}}]{morin_midM_08}
{Morin}, J., {Donati}, J., {Petit}, P., {et~al.} 2008, MNRAS, 390, 567

\bibitem[{{Morin} {et~al.}(2010){Morin}, {Donati}, {Petit}, {Delfosse},
  {Forveille}, \& {Jardine}}]{morin_lateM_10}
{Morin}, J., {Donati}, J., {Petit}, P., {et~al.} 2010, MNRAS, 1077

\bibitem[{{Moss}(2001)}]{moss_review_01}
{Moss}, D. 2001, in ASP Conference Series, Vol. 248, Magnetic fields across the
  Hertzsprung-Russell diagram, ed. S.~G.~Mathys \& D.~Wickramasinghe (San
  Francisco), 305

\bibitem[{{Mullan} \& {MacDonald}(2005)}]{mullan_macdonald_dynamo_05}
{Mullan}, D.~J. \& {MacDonald}, J. 2005, MNRAS, 356, 1139

\bibitem[{{Spruit}(2002)}]{spruit_dynamo_02}
{Spruit}, H. 2002, A\&A, 381, 923

\bibitem[{{Strassmeier}(2009)}]{strassmeier_review_09}
{Strassmeier}, K.~G. 2009, A. Ast. Rev., 17, 251

\bibitem[{{Tout} \& {Pringle}(1995)}]{tout_pringle_dynamo_95}
{Tout}, C.~A. \& {Pringle}, J.~E. 1995, MNRAS, 272, 528

\bibitem[{{Vidotto} {et~al.}(2010){Vidotto}, {Jardine}, {Opher}, {Donati}, \&
  {Gombosi}}]{vidotto_v374Peg_10}
{Vidotto}, A., {Jardine}, M., {Opher}, M., {Donati}, J.-F., \& {Gombosi}, T.~I.
  2010, MNRAS

\end{thebibliography}

\end{document}